\documentclass[aps,pre,twocolumn,showpacs,superscriptaddress,nofootinbib]{revtex4-1}  
\usepackage{graphicx}  
\usepackage{dcolumn}   
\usepackage{bm}        
\usepackage{amsmath, amsthm, amssymb}
\usepackage{bbold} 
\usepackage{mathrsfs}
\usepackage{array}
\usepackage[usenames]{color}
\usepackage{soul} 
\usepackage{ulem} 
\begin{document}

\normalem

\title{Unsupervised machine learning approaches to the $q$-state Potts model}
\author{Andrea Tirelli} 
\affiliation{International School for Advanced Studies (SISSA),
Via Bonomea 265, 34136 Trieste, Italy}
\author{Danyella O. Carvalho}
\affiliation{Departamento de F\'isica, Universidade Federal do Piau\'i, 64049-550 Teresina PI, Brazil}
\author{Lucas A. Oliveira}
\affiliation{Departamento de F\'isica, Universidade Federal do Piau\'i, 64049-550 Teresina PI, Brazil}
\affiliation{Instituto de F\'isica, Universidade Federal do Rio de
Janeiro Cx.P. 68.528, 21941-972 Rio de Janeiro RJ, Brazil}
\author{Jos\'e~P.~de Lima} 
\affiliation{Departamento de F\'isica, Universidade Federal do Piau\'i, 64049-550 Teresina PI, Brazil}
\author{Natanael C. Costa}
\affiliation{International School for Advanced Studies (SISSA),
Via Bonomea 265, 34136 Trieste, Italy}
\affiliation{Instituto de F\'isica, Universidade Federal do Rio de
Janeiro Cx.P. 68.528, 21941-972 Rio de Janeiro RJ, Brazil}
\author{Raimundo R. dos Santos} 
\affiliation{Instituto de F\'isica, Universidade Federal do Rio de
Janeiro Cx.P. 68.528, 21941-972 Rio de Janeiro RJ, Brazil}
\begin{abstract}
In this paper with study phase transitions of the $q$-state Potts model through a number of unsupervised machine learning techniques, namely Principal Component Analysis (PCA), $k$-means clustering, Uniform Manifold Approximation and Projection (UMAP), and Topological Data Analysis (TDA). Even though in all cases we are able to retrieve the correct critical temperatures $T_c(q)$, for $q=3,4$ and $5$, results show that non-linear methods as UMAP and TDA are less dependent on finite size effects, while still being able to distinguish between first and second order phase transitions. This study may be considered as a benchmark for the use of different unsupervised machine learning algorithms in the investigation of phase transitions. 
\end{abstract}


\maketitle

\section{\label{sec:intro}Introduction}
The study of phase transitions and critical phenomena is of crucial importance in Statistical Mechanics\,\cite{Stanley99,Herbut07}, with numerical simulations, and most notably Monte Carlo methods, playing a fundamental role in such analyses.
For most of the cases, these numerical approaches are performed at small lattices and exhibit strong finite-size effects, which, in turn, requires a careful analysis to avoid misleading results, in particular when dealing with phase transitions\,\cite{Landau21,Sandvik10,gubernatis16,Becca17}.
However, performing a finite-size scaling (FSS) analysis  \cite{Fisher71,Barber83} may be challenging and computationally expensive.
In view of this, devising auxiliary techniques to help in the identification of phase transitions, and which are capable of providing quantitatively accurate results for finite small system sizes, has become of crucial importance in this field.

Over the past years, with the advent of powerful machine learning (ML) techniques, this goal has become more reachable, with these methods being applied in many different research areas \cite{angra2017machine, shinde2018review, mater2019deep, akay2019deep}.
In the context of Condensed Matter and Statistical Physics there has been a great effort to develop supervised and unsupervised ML techniques to identify different phases (from their particular patterns), as well as to examine phase transitions\,\cite{Dunjko18,Carleo19a,Carrasquilla20}.
For instance, phase transitions have been examined through principal component analysis (PCA)\,\cite{Wang16,Hu17,Costa17a,Wang17a,Wang18a,Khatami19}, $t$-distributed Stochastic Neighbor Embedding ($t$-SNE)\,\cite{Wetzel17,Chng18,Zhang19a}, convolutional neural networks\,\cite{Chng17,Broecker17,Carrasquilla17,Kim18,Zhang18c,Khatami20}, restricted Boltzmann machines\,\cite{Nomura17,Huang17,Melko19}, intrinsic dimension analysis\,\cite{Mendes-Santos21a, Mendes-Santos21b}, and persistence homology\,\cite{donato2016,tran2020,Feng20,Olsthoorn20,Leykam21,Kashiwa21,tirelli2021}.
In particular, the use of persistence homology at the Topological data analysis seems a promising methodology to identify quantum critical points\,\cite{tirelli2021}.

Despite the advances, the use of ML methods in these fields is still in its infancy, with their validation and usefulness to different problems not being  entirely clear.
Here we contribute to this discussion by benchmarking a number of unsupervised machine learning techniques, using the classical $q$-state Potts model\,\cite{potts_1952,wu1982potts} as a testing ground.
This model describes interacting `classical spins' on a lattice, and has been used to capture the essential physics of a wide variety of systems, not necessarily magnetic ones\,\cite{wu1982potts}.
It is a generalization of the Ising model, in the sense that the spin variable on each lattice site can take on $q$ different values, instead of just two. 
Depending on the value of $q$, and of the spatial dimensionality of the systems, the Potts model may exhibit first or second order phase transitions, with challenging finite-size effects.
In view of these interesting features, we expect that the analysis of such a system will be crucial to assess performances and weaknesses of a range of unsupervised ML algorithms. 

In this work, we develop the following methods:
(i) the Principal Component Analysis (PCA)\,\cite{wold1987},
(ii) the $k$-means clustering\,\cite{macqueen1967some},
(iii) the Uniform Manifold Approximation and Projection (UMAP)\,\cite{mcinnes2018umap}, and
(iv) the Topological Data Analysis (TDA)\,\cite{Carlsson2009}.
We use such algorithms to examine the phase transitions of the two-dimensional Potts model for $q=3, 4$, and 5 -- from snapshots generated by Monte Carlo simulations--, presenting their results for different system sizes.
The paper is organized as follows: the model is presented in the next Section, while the ML methodologies are outlined in Section \ref{sec:methods}. Our results are presented and discussed in Section \ref{sec:results}, with final remarks and conclusions being left to Section \ref{sec:concl}.

\section{\label{sec:model} The $q$-state Potts model}

The Potts model describes interacting classical `spins' on a lattice, with each spin allowed to be in one of $q$ states\,\cite{wu1982potts}.
It may also be thought of as a classical unit vector, at a given site $\mathbf{i}$, pointing along the vertices of a polygon with $q$ sides, thus being oriented along a direction making an angle
\begin{equation}
\label{theta}
   \theta_i= \frac{2\pi n_i}{q},
\end{equation}
with some arbitrary direction, and where $n_i=0,1,2, ..., q-1$.
The Ising model corresponds to $q=2$, and, if one formally treats $q$ as a continuous parameter, the percolation problem is recovered in the limit $q\to 1$~\cite{wu1982potts}. 

The Potts Hamiltonian reads
 \begin{equation}\label{eq:ham-potts}
    \mathcal{H} = -J\sum_{\langle i,j \rangle}\delta_{\theta_i\theta_j} ,
\end{equation}
where the sum runs over a square lattice with periodic boundary conditions, $\langle i,j \rangle$ denotes nearest neighbour sites, and $J>0$ is the `exchange' coupling which sets the energy scale; $\delta_{\theta_i\theta_j}$ is the Kronecker delta function.
The Hamiltonian of Eq.\,\eqref{eq:ham-potts} is invariant under a discrete symmetry in which all spins are rotated by $2\pi/q$.

The Potts model has been extensively studied, and most of its properties are  well-known\,\cite{wu1982potts}. 
In particular, the critical temperature separating a ferromagnetic phase from a paramagnetic one on a square lattice is known from duality arguments to be \cite{potts_1952,wu1982potts} 
\begin{equation}\label{eq:crit-temp}
    T_c = \frac{J}{\ln(1+\sqrt{q})};
\end{equation}
throughout this paper we take a unit Boltzmann constant, $k_B=1$. 
It should also be noted that in two-dimensional systems the transition is continuous (second order) if $q\leq 4$, with the critical exponents being known exactly \cite{denNijs79}, while for $q>4$ the transition is discontinuous (first-order)\,\cite{wu1982potts}.
In view of this wealth of exact data, the Potts model may be used as a testing ground for new methodologies, as in the case of this work, for unsupervised machine learning techniques.

\section{\label{sec:methods}Methods}
In this section we briefly describe the Machine Learning algorithms used in the analysis of phase transitions of the Potts model: all these techniques are $unsupervised$ methods, i.e.\,methods working on unlabelled datasets, where  typical goals are to automatically find patterns within such data, or to reduce their dimensionality without loss of information. 
We resort to four methods. 
The first two are the Principal Component Analysis and the $k$-means clustering, standard and well-known ML approaches which have been recently adapted to solve problems in Statistical Physics and Condensed Matter \cite{Costa17a,Hu17,Wetzel17,Wang16,Wang18a,Carrasquilla17}. 
The third one is the Uniform Manifold Approximation and Projection (UMAP) algorithm~\cite{mcinnes2018umap}:
a manifold learning technique for non-linear dimensionality reduction. 
In a way, using UMAP corresponds to using PCA, but with non-linear, topology-oriented settings. 
Finally, we examine the Potts model through a Topological Data Analysis (TDA), which is a new class of techniques based on Applied Computational Topology \cite{Carlsson2009}, and has been recently used in the study of phase transitions \cite{Olsthoorn20, Leykam21,tran2020, tirelli2021}. 
In what follows, we present highlights of these four methods.

\subsection{Principal Components Analysis}
\label{subsec:PCA}

The Principal Component Analysis (PCA)\,\cite{wold1987,abdi2010,ringner2008principal} is a {\it linear} dimensionality-reduction technique whose primary scope is to reduce the dimensionality of large data sets, by transforming the (often large) set of covariates into a smaller one, while still retaining most of the information in the original set of features. 
Although the reduction in the number of variables in a data set usually comes at the expense of accuracy, i.e.\ loss of information, one may perform such a dimensionality reduction by rendering information loss as small as possible. This kind of method is often used as a preprocessing technique, with the reduced data set created by PCA being used to feed supervised machine learning approaches. 
In this work, the reduced data is processed through a clustering method instead, as discussed below.

Let $X$ describe a dataset, i.e.~a matrix with $n$ rows (samples) and $m$ columns (features); we denote by $X_i$ the $i$-th column of $X$. Then, PCA works through a number of computational steps as follows: 
\begin{itemize}
    \item {\it standardization}: each variable is properly scaled so that each contributes equally to the analysis. This corresponds to defining a new dataset $X^{\mathrm{std}}$, such that, 
    \[
    X_i^{\mathrm{std}} = \frac{X_i - \mu_i}{\sigma_i},
    \]
    where $\mu_i$ and $\sigma_i$ are the mean and the standard deviation of the vector $X_i$, respectively. The operations of subtraction and division are performed component-wise;

\item {\it covariance matrix computation}: this step is needed to obtain the correlations between the features of $X$. The covariance matrix $\mathrm{Cov}(X)$ is defined as 
\begin{align}
 \mathrm{Cov}(X)_{i, j} &= \mathrm{cov}(X_i, X_i) \nonumber\\
 &=\mathbb{E}((X_i -\mathbb{E}(X_i))((X_j -\mathbb{E}(X_j)),
\end{align}
where $\mathbb{E}$ stands for the mean value operation;
\item {\it computation of the spectrum of $\mathrm{Cov}(X)$}: eigenvectors of the covariance matrix are the principal components and are derived from the initial features by performing linear combination. 
In a way, this can be interpreted as a {\it basis change} in the features space. The most important characteristics of these new features is that they are uncorrelated and most of the information within the initial variables is squeezed or compressed into the first components. 
\end{itemize}
Usually, most of the information of $X$ is contained in a few principal components, namely the ones whose corresponding eigenvalues have the highest modulus. As such, a new dataset $\bar{X}$ can be constructed as follows: the $i$-th column $\bar{X}_i$ of $\bar{X}$ is the $i$-th eigenvector of $\mathrm{Cov}(X)$, assuming that eigenvectors have been sorted in ascending order, according to the magnitude of their eigenvalues. 
In most cases, such as those discussed here, one only selects  the principal components whose eigenvalues are larger than a certain threshold: in this way we significantly reduce the dimensionality of the dataset, while retaining most of its information content. We note that, while being a very commonly used technique, PCA has some limitations: for example, it fails to detect {\it non-linear} correlations between features.

\subsection{$k$-means clustering}
\label{subsec:kmeans}

The $k$-means method\,\cite{macqueen1967some, steinhaus1956division} is an algorithm that systematically divides the dataset within a predefined number of clusters.
That is, given a dataset, $X$, composed of $n$ observations, its aim is to partition such samples into $k$ clusters, such that each observation belongs to the cluster with the nearest mean (cluster centres or cluster centroid), which serves as a prototype of the cluster. The way in which $k$-means performs this task is by minimizing the within-cluster variances. 

More formally, given a dataset $X$, i.e. a matrix $X$ of dimension $(n, m)$, let $x_i$ denote the $i$-th row, for $i=1, \dots n$. $k$-means aims to partition the samples $x_1, \dots, x_n$ into $k$ $(k\leq n)$ clusters $C=\{C_1,\dots, C_k\}$ by optimizing the within-cluster sum of squares: therefore, the minimization problem can be written as
\[
\underset{C}{\mathrm{argmin}}\sum_{i=1}^k\sum_{x\in C_i}||x-\mu_i||^{2},
\]
where $|| \cdot||$ denotes the Euclidean vector norm and $\mu_i$ is the mean of points belonging to the cluster $C_i$; such a mean is often called {\it centroid} of the cluster. Since solving exactly such a minimization problem is NP-hard\,\cite{MAHAJAN201213}, heuristic techniques are commonly used to obtain approximate solutions in polynomial time. One such technique is the following iterative algorithm: 
\begin{itemize}
    \item we start with an initial (random) assignment of the $k$ centroids $m_1^1, \dots, m_k^1$;
    \item we proceed by alternating an assignment step and an update step, until a certain stopping criterion is met.
\end{itemize}
In the assignment step, each observation is assigned to the cluster corresponding to the nearest centroid, with respect to the euclidean distance, so that the $i$-th cluster at iteration $t$ can be written as 
\[
C_i^t = \{x: || x - m_i^t|| \leq || x - m_i^t||, \forall j, 1 \leq j \leq k\}.
\]
In the update step, the centroids are recomputed, with respect to the new partition into clusters generated by the assignment step, 
\[
m_{i}^{t+1}=\frac {1}{\left|S_{i}^{t}\right|}\sum _{x_{j}\in S_{i}^{t}}x_{j}
\]
The algorithm has converged when the assignments no longer change. Although the algorithm is not guaranteed to find the optimum\,\cite{hartigan1979algorithm}, it normally produces an adequate partition of the dataset into clusters. 

\subsection{Combining PCA and $k$-means}
\label{subsec:PCAandkmeans}

Even though PCA and $k$-means perform two seemingly unrelated tasks, they are often combined during the unsupervised analysis of a dataset. In particular, given a dataset $X$, which might have a large number of samples and features, one first uses PCA to reduce the dimensionality, producing a new dataset $\bar{X}$, with less features, but retaining (most of) the information content of $X$. Cluster analysis is then performed on $\bar{X}$ via $k$-means. Stacking together PCA and $k$-means in this way is a common practice, especially when the number of principal components is 2 or 3: in this case, samples on the initial dataset $X$ can be visualized as points in the two-dimensional plane or three-dimensional space, and clusters can be represented as coloured point clouds. This approach is the one adopted in the present work. 

\subsection{UMAP}
\label{subsec:UMAP}

The UMAP is a non-linear dimensionality reduction technique, which first appeared in Ref.\,\cite{mcinnes2018umap}.
Due to the mathematical complexities behind the method, whose foundations are based on advanced topics at the intersection between Riemannian Geometry and Algebraic Topology, here we skip the details about its implementation; see, e.g., \cite{sainburg2020parametric}, for a pedagogical introduction.
The UMAP has proved to be a very efficient and scalable algorithm, with improved performances with respect to analogous non-linear dimensionality reduction techniques, such as $t$-SNE\,\cite{van2008visualizing}. 

In broad terms, the UMAP algorithm falls into the category of dimension-reduction techniques: UMAP performs such task through a procedure called \textit{manifold learning}. The idea behind manifold learning is that data points embedded in a higher dimensional space tend to distribute on regular lower dimensional subspaces, whose nature is determined by the common features of the data points.
For instance, Monte Carlo snaphots for the Potts model at a given temperature $T$ will gather along a lower dimensional submanifold in the high dimensional space of all possible Monte Carlo snapshots. Therefore, the aim of UMAP is to map points in a high dimensional space to points a lower dimensional space in a faithful way, i.e.\ by preserving their local and global structures. 

Let $X$ be a point-cloud, and $\tilde{X}$ the non-linear projection of $X$ to the feature space constructed by UMAP. The key point is that UMAP is capable of preserving as much as possible the global structure -- topology -- of $X$ in the projection $\tilde{X}$. At its core, UMAP constructs a high-dimensional graph representation of the data and then optimizes a corresponding low-dimensional graph representation, making it as structurally similar as possible to the higher dimensional one. The UMAP algorithm depends on several hyper-parameters: the most important one is $n_\text{neigh}$, the number of approximate nearest neighbours used to construct the initial high-dimensional graph. Indeed, it controls how UMAP balances local versus global structure: low values pushes UMAP to focus more on local structure, by constraining the number of neighbouring points considered when analysing the data in high dimensions, while high values lead UMAP towards representing the big-picture structure, losing fine detail. In our case, we are interested in observing differences in the global structure of the data, for variations of the order parameter for the phase transition, so we shall use UMAP with a large value of $n_\text{neigh}$.

\subsection{Topological Data Analysis}
\label{subsec:TDA}

The Topological Data Analysis is a set of methods whose aim is to analyse datasets from the point of view of topology. 
In this work, we follow the pipeline devised in Ref.\,\cite{tirelli2021}, where one may find a detailed explanation of the many steps for implementation and analysis.

Briefly speaking, the TDA uses computational techniques from Applied Topology to describe a given dataset by a group of topological features, called persistence diagrams,
which are computed by Persistence Homology\,\cite{edelsbrunner2010computational}.
That is, from a dataset $X$, interpreted as a point cloud in a higher-dimensional space, one is able to obtain a set of topological features $\mathcal{D}(X)$ which depends on certain parameters, modelling the complexity of the topological invariants that one wishes to extract from $X$.
Therefore, fixing such a set of topological invariants $\mathcal{I}$, one can build a map
\begin{equation}\label{top_emb}
X \longrightarrow \mathcal{D}_X,    
\end{equation}
by associating a point cloud $X$ to its persistence diagram $\mathcal{D}_X$.
Such a mapping, also called {\it topological embedding}, allows us to compare two different point clouds $X_1$ and $X_2$ by endowing the space $\mathcal{PD}$ of all persistence diagrams with a metric space structure (Wasserstein, Betti or Bottleneck distances) between the diagrams.

\begin{figure}[t]
	\centering
	\includegraphics[scale=0.30]{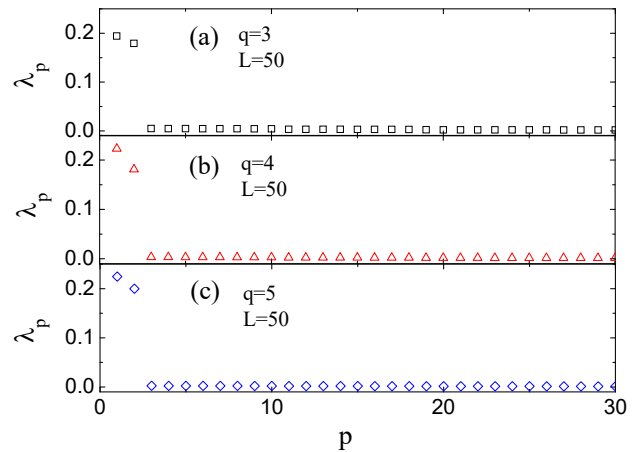}
	\caption{The PCA normalized eigenvalues for the Potts model for (a) $q=3$, (b) $q=4$, and(c) $q=5$.}
	\label{fig:pc_components}
\end{figure}

Through persistence diagrams we compare different point clouds in terms of their topological similarity, which in turn allows us to set up an algorithm separating point clouds into clusters.
In other words, given $n$ point clouds, $X_1, \dots, X_n$ (for the purposes of this work, one can assume $X_i$ as a snapshot of spin configurations at a given temperature obtained through a Monte Carlo simulation), one can construct a square matrix $M_X$ of dimension $n$, $M_X=(m_{ij})$, by letting 
\begin{equation}\label{eq:dist_matrix}
m_{ij} = \tilde{d}(X_i, X_j),
\end{equation}
where $\tilde{d}$ is the distance measure constructed with persistence diagrams.
The key point is that the more topologically similar the point clouds $X_i$ and $X_j$ are, the smaller the value $\tilde{d}(X_i, X_j)$ is. As such, $M_X$ can be used, by means of \textit{clustering algorithms}, to group together point clouds sharing the same topological features, and separate those with an inherently different topology. 
As described Ref.\,\cite{tirelli2021}, this is performed by using fuzzy spectral clustering\,\cite{von2007, liu2018, jimenez2008fuzzy} for the identification of two clusters (one corresponding to point clouds below the critical temperature, the other for point clouds above the critical temperature).
It leads to the construction of a membership degree function
\[
l=(l_0, l_1): \{X_1, \dots, X_n\} \rightarrow [0, 1]^2, 
\]
such that $l_0(X_i)$ is the membership degree to the first cluster of the point cloud $X_i$, and $l_1(X_i)$ is the membership degree to the second cluster of the point cloud $X_i$; $l_0(X_i)+l_1(X_i)=1$ for all $i$.
Then, the critical point is obtained by analyzing the sequence 
\begin{equation}\label{eq:member_list} 
\bar{l} = (l_0(X_1), \dots, l_0(X_n)),
\end{equation}
being one of the endpoints (based on the position) of the longest quasi-constant subsequence of $\bar{l}$.
It is worth noticing that this sequence can be segmented in smaller subsequences, based on the range of variation of the membership degree, and that the critical point could be detected by a simple linear time algorithm scanning $\bar{l}$. 

\begin{figure}[t]
	\centering
	\includegraphics[scale=0.35]{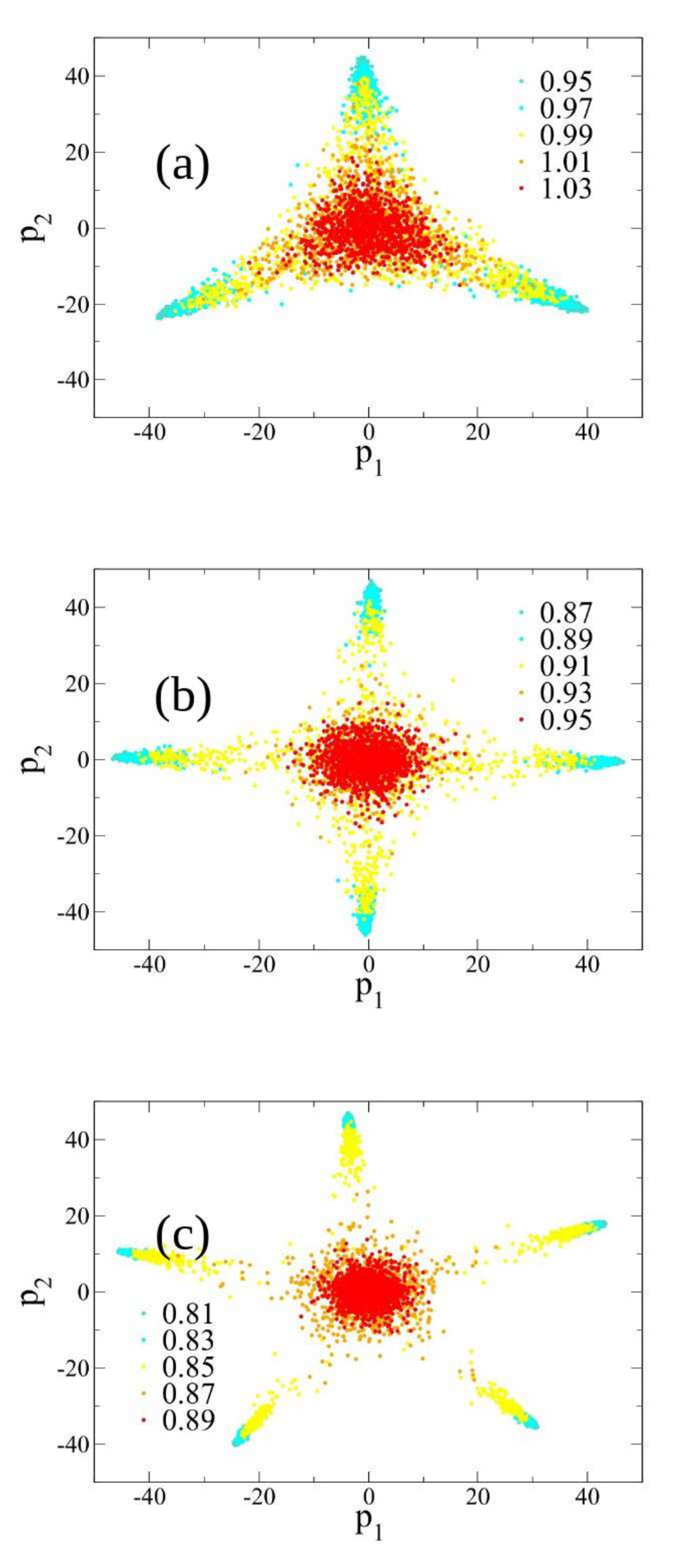}
	\caption{Cluster division through the PCA algorithm around the critical point, for (a) $q=3$, (b) $q=4$, and(c) $q=5$.
	In each panel a colour is ascribed to each temperature, as shown.}
	\label{fig:kmeans}
\end{figure}

\section{Results}\label{sec:results}

We now apply these ideas to the classical \textit{q}-state Potts model, Eq.\,\eqref{eq:ham-potts}, on $L\times L$ square lattices with periodic boundary conditions.
We examine this Hamiltonian by standard Monte Carlo (MC) methods, feeding the Machine Learning approaches with snapshots of the spin configurations throughout the simulations.
These snapshots are stored as vectors, $\vec{\sigma} ~=~ (n_1, n_2, \cdots, n_{L^2})$, of length $L^2$, with $n_i = 0, 1, \cdots, q-1$.
Through MC sampling, by skipping many sweeps to guarantee independent measurements, we generate $m$ configurations at a given temperature, for a total of $l_T$ different temperatures.
Typically, in this work, we stored $m \approx 500$ spin configurations, for $l_T \approx 20-30$ temperatures around the expected critical point.
Therefore, the resulting dataset, $X_q$, is a matrix with dimension $m\,l_T \times L^2$, treating a single configuration as a row vector.
In order to assess effects arising from the finite size of the lattices, we consider $L=20, 30, 40, 50, 60,$ and 80.

\subsection{Phase transitions via PCA}

Here, we perform the PCA approach to investigate the phase transitions.
By following the procedures outlined in Sec.\,\ref{subsec:PCA}, we use the matrix $X_q$ to feed the PCA algorithm, therefore selecting the leading components.\footnote{At this point, we note that when feeding PCA, instead of providing $\vec{\sigma}= [n_1, n_2, \cdots ]$, we provide their directions, i.e.~$\vec{\sigma}= [(\cos \theta_1, \sin \theta_1), (\cos \theta_2, \sin \theta_2), \cdots ]$. Thus, our dataset matrix has $2\times L^2$ columns. We emphasize that such a change does not affect the final results, just the spacial position (patterns) of the clusters in PCA space.}
This allow us to obtain a new dataset $\tilde{X_q}$ of dimension $m\,l_T\times p$, with $p \ll L^2$, i.e. we embed the initial data points of $X_q$ in the {\it PCA space}.
In Figure \ref{fig:pc_components}, we present the first 30 normalized eigenvalues, $\lambda_{p}$, for $q=$3, 4, and 5. 
Notice that, for all cases, the first two eigenvalues are orders of magnitude larger than the others.
Therefore, when choosing the dimension of the PCA space, we may set $p=2$, since the first two components contain most of the information of $X_q$.

Denoting by $p_1$ and $p_2$ the first two principal components,
Fig.\,\ref{fig:kmeans} presents the dataset projection in this two-dimensional PCA space.
It is important to note that the number of clusters at low temperatures is  exactly equal to the number of degenerate ground states, $q$, each of which corresponding to a state in which the symmetry of the Hamiltonian is broken. 
This further confirms the conjecture that the low-temperature projection in the PCA space provides information about the intrinsic symmetries of Hamiltonian.  
In addition, the change from a single-cluster to many-clusters occurs near the exact critical temperatures given by Eq.\,\eqref{eq:crit-temp}, namely $T_c = 0.9950$, 0.9102, and 0.8515, for $q=3$, 4, and 5, respectively.

\begin{figure}[t]
	\centering
	\includegraphics[scale=0.30]{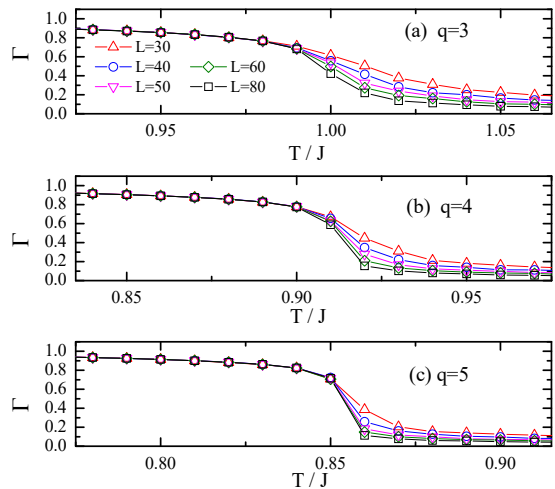}
	\caption{The PCA order parameter, $\Gamma$, for (a) $q=3$, (b) $q=4$, and(c) $q=5$, for different linear lattice sizes, $L$.}
	\label{fig:gamma}
\end{figure}

\begin{figure}[t]
	\centering
	\includegraphics[scale=0.30]{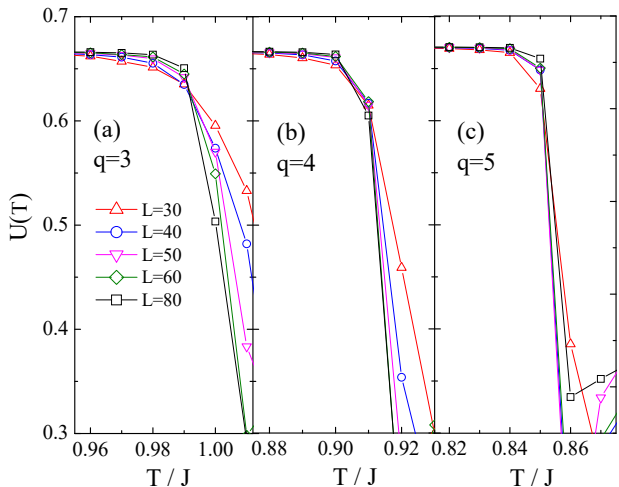}
	\caption{The PCA ``Binder cumulant'', $U(T)$, for (a) $q=3$, (b) $q=4$, and(c) $q=5$. }
	\label{fig:binder}
\end{figure}

In order to further investigate $T_c$, we should define a PCA analogue of the order parameter.
To this end, we denote by $p_i^x$, $i=1, 2$ the two-dimensional projection of a given snapshot $x$, at the corresponding principal components.
Since the eigenvalues $\lambda_1$ and $\lambda_2$ have the same order of magnitude, it is convenient to compute
\[
\gamma_x = \sqrt{||p_1^x|| ^2+ ||p_2^x||^2},
\]
which, when averaged over all samples at a fixed temperature, $T$, leads us to our definition of the PCA order parameter,
\begin{equation}
\Gamma(T) = \frac{1}{m L} \sum_{x=1}^{m} \gamma_x (T) = \bigg\langle \frac{ \gamma_x }{L} \bigg\rangle_T.
\end{equation}
The behavior of $\Gamma$ as a function of temperature is displayed in Fig.\,\ref{fig:gamma}, and indeed resembles that of an actual order parameter for the system.
That is, $\Gamma$ decreases as a function of $T$, and its steepness of the descent increases around the transition temperature. Even though this behaviour is consistent for all values of $L$, it is much more evident for larger lattice sizes, emphasizing the finite size effects in the PCA approach. 
Due to the similarities of $\Gamma (T)$ with the magnetization, one may naturally consider the PCA analogue of the Binder cumulant\,\cite{binder1981finite, binder1981critical} in terms of $\Gamma$.
Accordingly, we define 
\begin{equation}
U(T) = 1 - \frac{\big\langle \big( \frac{ \gamma_x }{L} \big)^{4} \big\rangle_T}{3\big\langle \big( \frac{ \gamma_x }{L} \big)^{2} \big\rangle_{T}^{2}},
\end{equation}
whose temperature dependence is displayed in Fig.\,\ref{fig:binder}, for different lattice sizes.
Interestingly, $U(T)$ seems to cross around the critical points, as expected for the actual Binder cumulants.
In fact, the behaviour of $U(T)$ for $q=5$ is very noisy, although a significant change in the quantity is observed around its $T_c$.

\begin{figure}[t]
	\centering
	\includegraphics[scale=0.29]{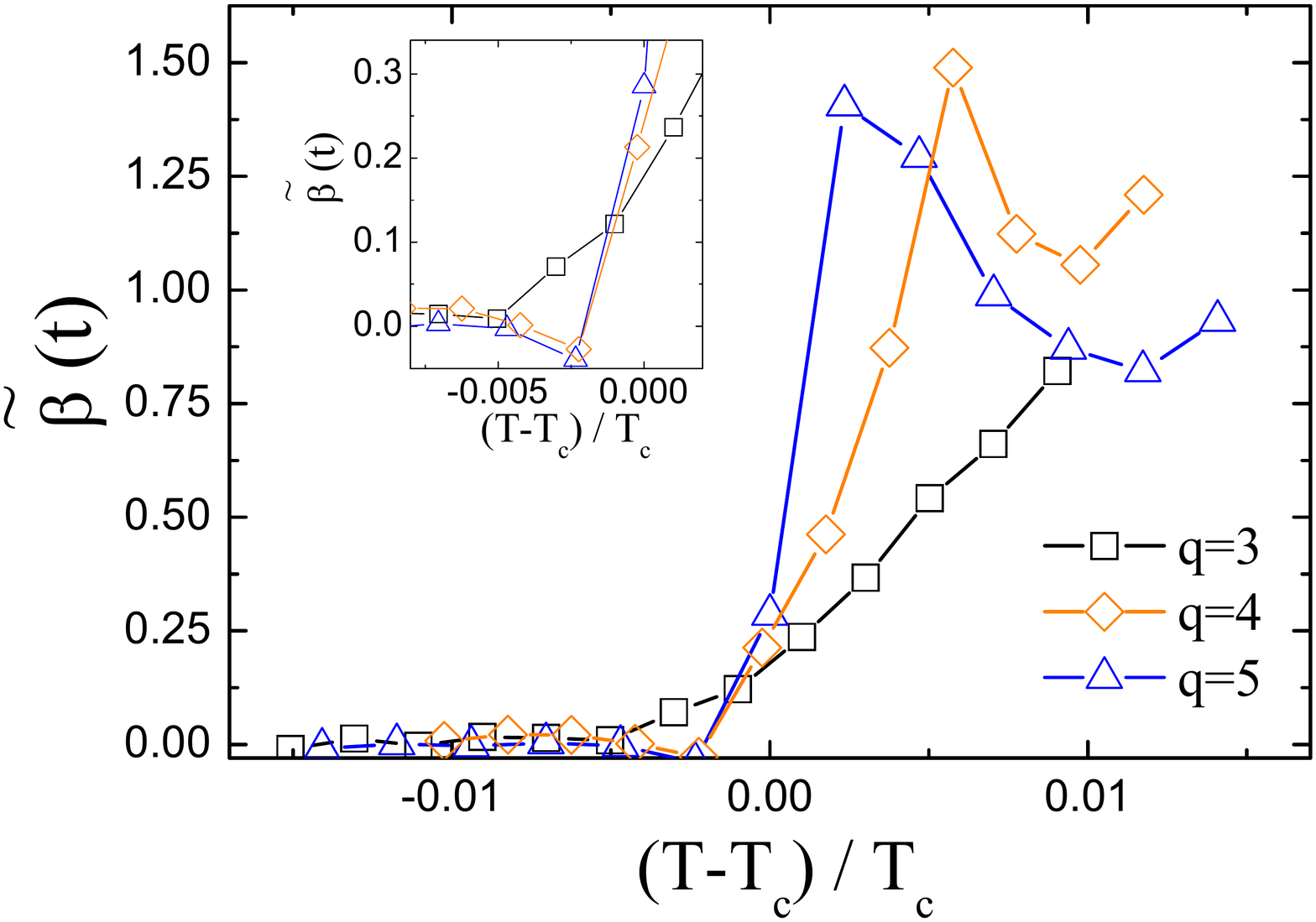}
	\caption{The `effective' critical exponent $\tilde{\beta}$, obtained through Eq.\,\eqref{eq:exponents}, for $q=3$, 4 and 5, as functions of the reduced temperature, $t \equiv (T-T_c)/T_c$. Data have been collected using $L_1=60$ and $L_2=80$.}
	\label{fig:exponents}
\end{figure}

The crossings of $U(T)$ around the critical temperature strongly suggest that, to some extent, the well-known FSS theory \cite{Fisher71,Barber83} is still valid in this PCA space; that is, close to the critical temperature and for $L\gg1$, one should have 
\begin{equation}
    \Gamma_{L}(T) = L^{-\beta/\nu} f\big[ L^{1/\nu} (T-T_c) \big],
    \label{eq:FSS}
\end{equation}
where $f[z]$ is a scaling function in terms of the variable $z$, and $\beta$ and $\nu$ are the critical exponents for the order parameter and for the correlation length, $\xi\sim |T-T_c|^{-\nu}$, respectively. 
Further, Eq.\,\eqref{eq:FSS} establishes that $L^{\beta/\nu}\Gamma_L(T)$ approaches a constant value, independent of $L$, as $T\to ~T_c$, which may be used to extract critical properties in a systematic way \cite{dosSantos81a}.
In particular, by considering two systems of different linear sizes, $L_1<L_2$, one may define 
\begin{align}\label{eq:exponents}
    \tilde{\beta}(t) \equiv  -\frac{\ln\big[\Gamma_{L_2}(t)/\Gamma_{L_1}(t)\big]}{\ln (L_2/L_1)},
\end{align}
where $t \equiv (T-T_c)/T_c$. 
Let us then examine the evolution of $\tilde{\beta}(t)$ with $q$, and check how the presence of a discontinuous transition for $q>4$ manifests itself in the data, as plotted in Fig.\,\ref{fig:exponents}. 
Starting with $q=3$, we see that $\tilde{\beta}\approx 0.17(5)$, and hardly changes in the $t\lesssim 0$ region, consistent with the exact value \cite{denNijs79} of $\beta/\nu = 2/15$; it then rises steadily beyond $t=0$. 
Note that it is $T_c$ for the infinite system which enters Eq.\,\eqref{eq:FSS}, so that this way of estimating $\tilde{\beta}$ would still demand some adjustments to account for these additional finite-size corrections \cite{dosSantos81a}; nonetheless, for our purposes here, this simplified version suffices to highlight the trend. For $q=4$, contrasting with the constant behaviour observed for $t\lesssim0$ when $q=3$, we see that $\tilde{\beta}$ drops slightly near $t=0$ and then rises sharply as $t$ increases. 
The behaviour for $q=5$ follows along similar lines, but with a steeper rise at $t\gtrsim0$ than for $q=4$; this can be attributed to the fact that for $q=5$ the transition is already discontinuous, while the case $q=4$ may be considered as marginal. 
For completeness, we should mention that a first order phase transition is consistent with $\beta/\nu = 0$ \cite{Fisher82}, a behaviour which should set in at much larger lattices.
Thus, when analysed in conjunction with a solid framework, such as FSS theory, the presence of a first order transition should manifest in PCA data as a sharp increase in $\tilde{\beta}$ near $t=0$.  

\begin{figure}[t]
	\centering
	\includegraphics[scale=0.28]{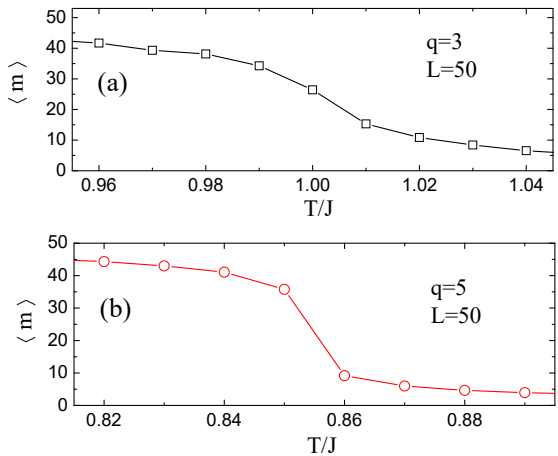}
	\caption{Average distance from the origin of the $k$-centroids of the PCA point-clouds as a function of the temperature, for (a) $q=3$, and (b) $q=5$. Data are from a lattice with linear size $L=50$. Notice the small scale of the temperature axes.}
	\label{fig:avecent}
\end{figure}

\subsection{Phase transitions via PCA + \textit{k}-means}

We now turn to the analysis of the Potts model by combining $k$-means with PCA.
That is, from the PCA projection onto the reduced space, we perform the $k$-means algorithm on the new dataset $\tilde{X_q}$, thus refining the clustering analysis.

Figure \ref{fig:kmeans} shows that $q$ clusters ar formed at low-temperatures, so that for our $k$-means analysis we divide the dataset into $C_1, \dots, C_{k=q}$ clusters, and we let $m_1, \dots, m_{k=q}$ be their corresponding centroids. 
Such a division leads us to define the PCA+$k$-means order parameter as the average distance of the centroids from the origin, i.e.\ $\langle m \rangle=(1/k)\sum_{i=1}^k ||m_i||$.
Figure \ref{fig:avecent} shows $\langle m \rangle$ as a function of temperature for $q=3$ and 5, both for a lattice with linear size $L=50$. 
The behaviour of $\langle m\rangle$ is analogous to that of $\Gamma$ (see Figure \ref{fig:gamma}), showing a steep decrease around the exact critical temperature, as given by Eq.\,\eqref{eq:crit-temp}; thus, similarly to $\Gamma$, $\langle m\rangle$ also tracks the behaviour of the order parameter for the Potts transition.

\begin{figure}[t]
	\centering
	\includegraphics[scale=0.28]{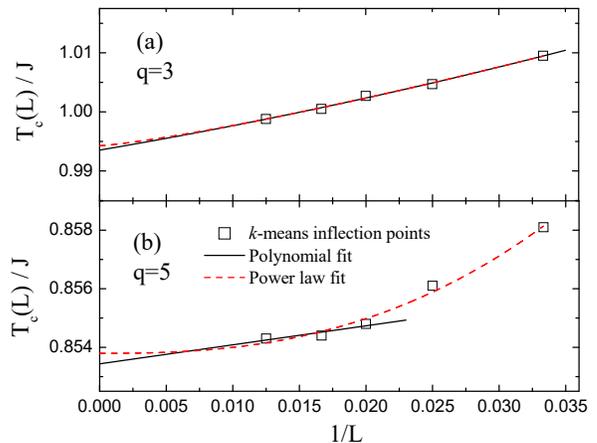}
	\caption{Scaling analysis of the predicted transition points combining PCA and $k$-means, for (a) $q=3$, and (b) $q=5$. Solid black curves are polynomial fittings, while red dashed lines are power law ones.}
	\label{fig:scaling}
\end{figure}

Pushing this idea further, we interpolate the data points assuming a fit to a continuous nonlinear function, in terms of which we may define the (pseudo) critical temperature, $T_c$, for that lattice size, as the inflection point of the curve.
By following the same procedure for different sizes, $L$, we obtain a sequence of points, $T_c(L)$, which is plotted in Fig.\,\ref{fig:scaling} in terms of $1/L$ to highlight the extrapolation towards $L\to\infty$. 
We note that although the size dependence of the critical temperature is rather weak, one can obtain more accurate estimates for $T_c$ by performing polynomial and power law extrapolations of $T_c(1/L)$ for $1/L\to0$: the extrapolated values are in excellent agreement with the exact results from Eq.\,\eqref{eq:crit-temp}.

\begin{figure}[t]
	\centering
	\includegraphics[scale=0.6]{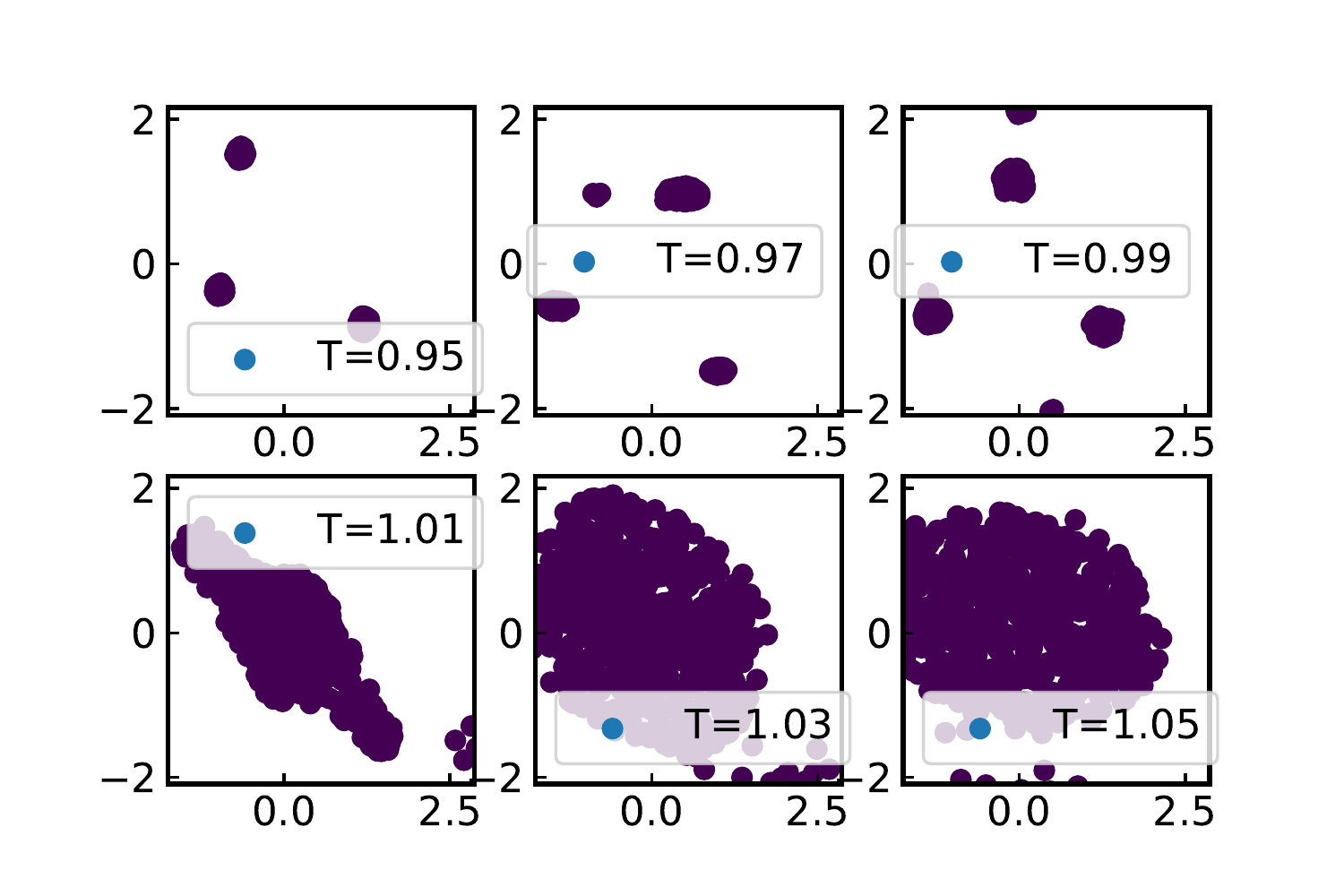}
	\caption{UMAP projections onto a two-dimensional latent space, for $q=3$ and $L=40$ and different values of the temperatures, around the critical one,  $T_c=0.9950$.}
	\label{fig:umapq3}
	\centering
	\includegraphics[scale=0.6]{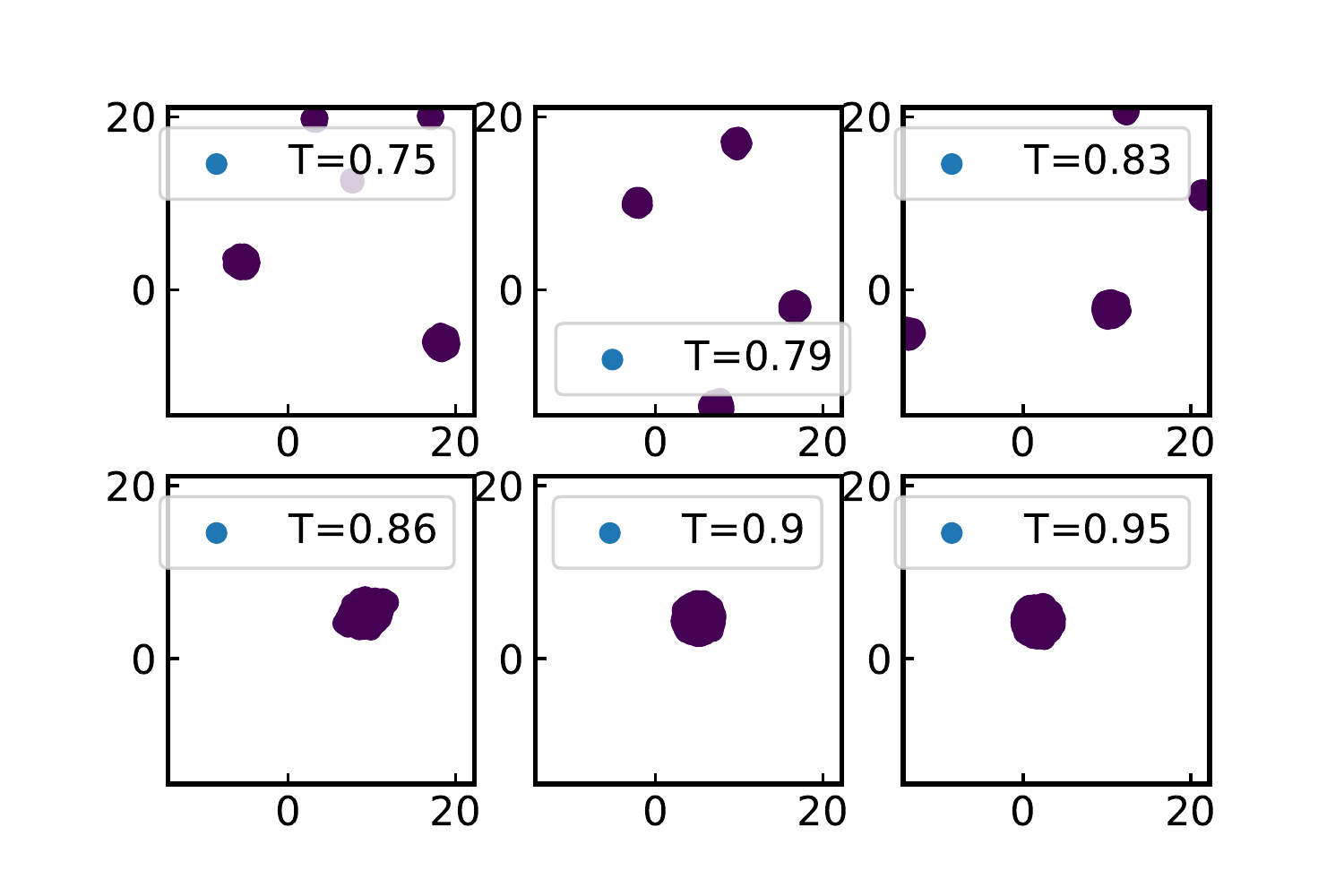}
	\caption{Same as Fig.\,\ref{fig:umapq3}, but for $q=5$, in which case $T_c=0.8515$.}
	\label{fig:umapq5}
\end{figure}

\subsection{Phase transitions via UMAP}

The analysis we perform with the UMAP algorithm is very similar to the one using PCA reduction. Essentially, we project the dataset $X_q$ onto a lower dimensional space and investigate how variations of the order parameter around the transition point affect the arrangement of data points in the space. 

Figures \ref{fig:umapq3} and \ref{fig:umapq5} show the dimensionality reduction for $q=3$ and $q=5$, respectively.
In each case one sees that for temperatures below $T_c$ the data-points form well-separated clusters, whereas for $T>T_c$  the corresponding Monte Carlo snapshots are all gathered in a single cluster.
Even though this behaviour is quite similar to those of PCA and $k$-means discussed in the previous section, we notice that with UMAP the division into clusters for $T<T_c$ is more uniform; the same holds for the concentration in a single cloud for $T>T_c$. This shows the power of the nonlinear analysis performed on UMAP over the linear ones in PCA.
In other words, clustering analysis through UMAP already provides a good estimate for the critical temperature by itself, without resorting to analogies with Binder cumulants, as done for the PCA case.


\subsection{Phase transitions via TDA}

The previous analyses (UMAP and PCA) emphasize the idea that the \textit{topology} of the dataset is a crucial ingredient to study phase transitions through machine learning methods.
In view of this, one may expect that the use of TDA in the context of the Potts model should lead to even better results.

\begin{figure}[t]
	\centering 
	\includegraphics[scale=0.65]{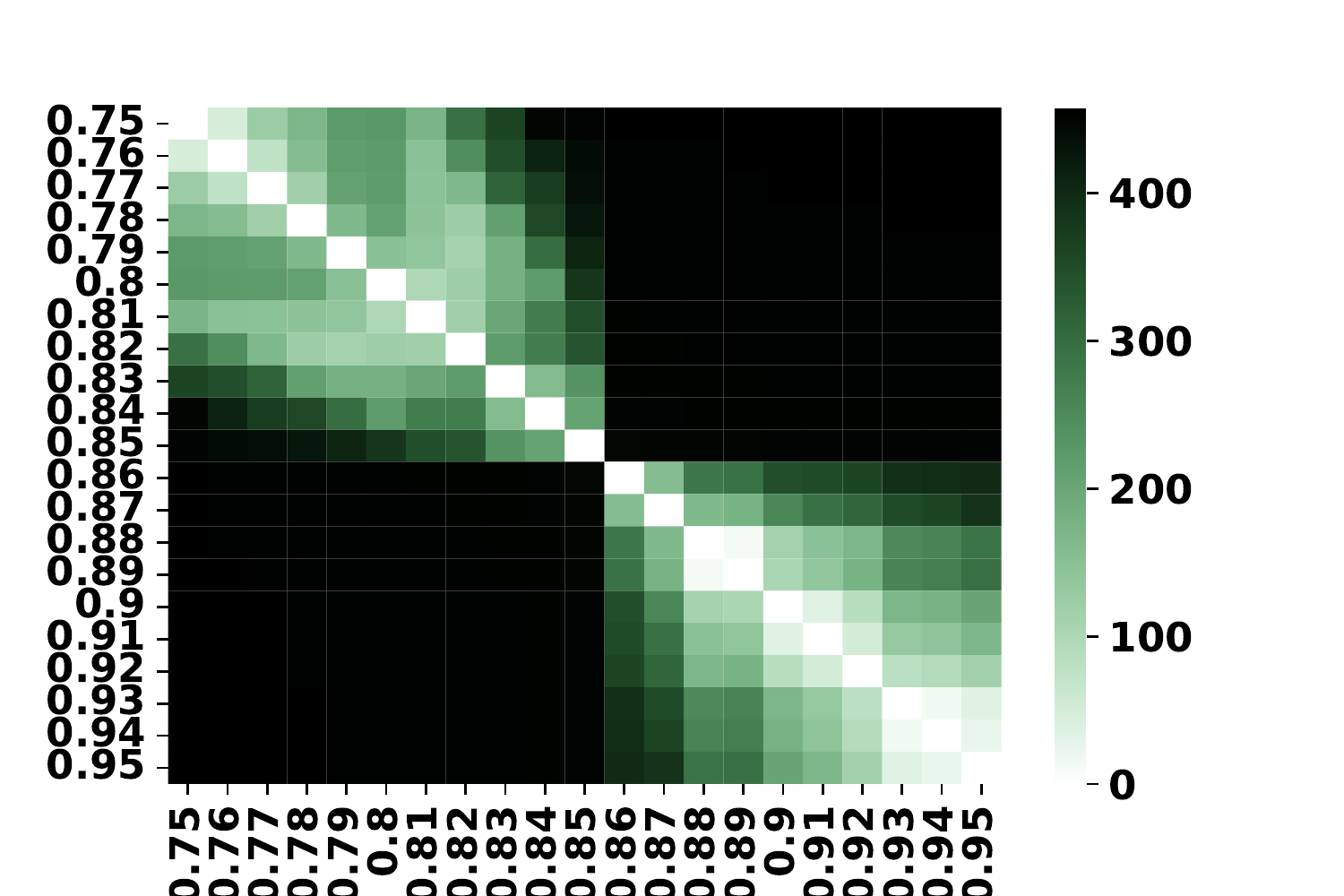}
	\caption{Heat-map representing the distance matrix $M$ for the 5-state Potts model on an $80\times80$ lattice, using the Betti distance with $p=2$; column labels are the different values of the temperature, $T$.}
	\label{fig:heatmap}
\end{figure} 

We proceed by following the pipeline outlined in Ref.\,\cite{tirelli2021}: first, using TDA we project data-points into the persistence space and then, with fuzzy spectral clustering, we assign a membership-score to each point-cloud at a given temperature. 
For instance, Fig.\,\ref{fig:heatmap} exhibits a heat-map representation of the distance matrix constructed from persistence diagrams, Eq.\,\eqref{eq:dist_matrix}, for the case $q=5$ and $L=80$.
One can identify two darker clusters in the bottom-left and upper-right zones, as well as two lighter clusters in the upper-left and bottom-right zones.
Such pattern conveys the following information: (1) point-clouds corresponding to temperatures $T$ ranging from $0.75$ to $0.85$ are all topologically similar, and the same holds for the point clouds in the range $0.86$-$0.95$;
(2) point clouds corresponding to $0.75 \leq T\leq 0.85$ are topologically different from all point-clouds in the range $[0.86, 0.95]$, and vice-versa.
This difference in topological invariants at different temperature ranges is a clear evidence of the phase transition.
Further, the fact that this change occurs very close to the exact critical temperature, namely $T_c=0.8515$, can hardly be regarded as fortuitous. 


\begin{figure}[t]
	\centering
	\includegraphics[scale=0.44]{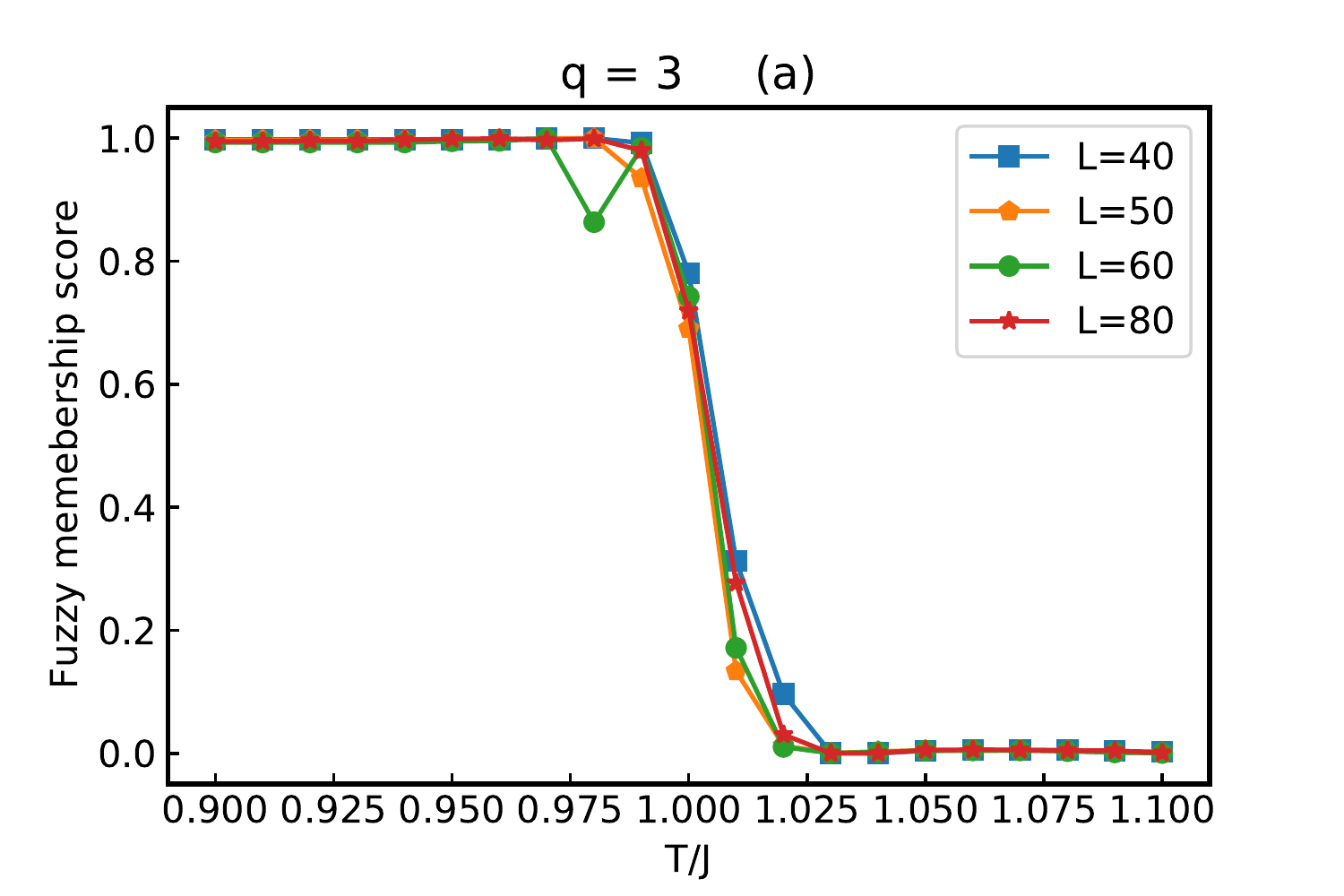}
	\includegraphics[scale=0.44]{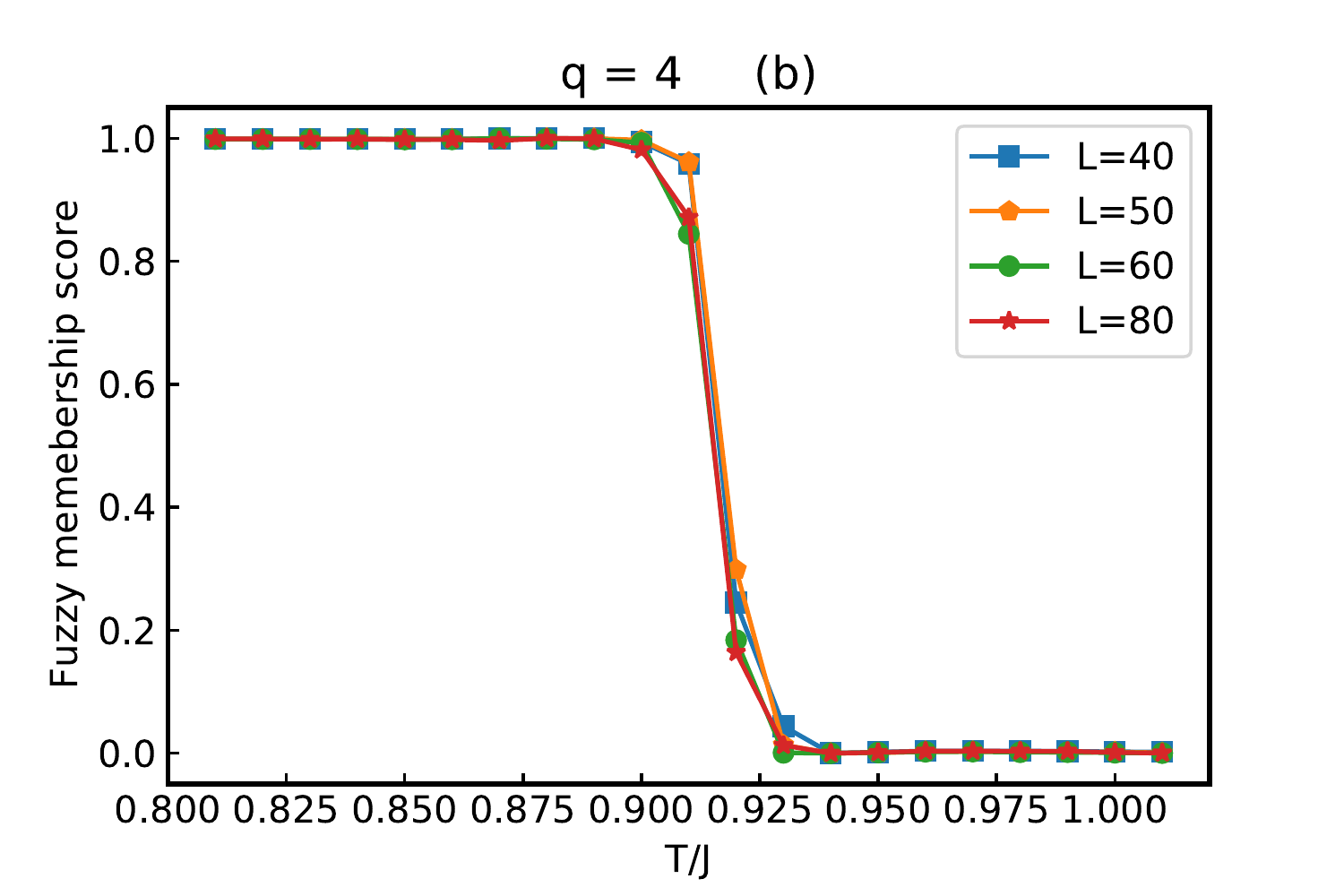}
	\includegraphics[scale=0.44]{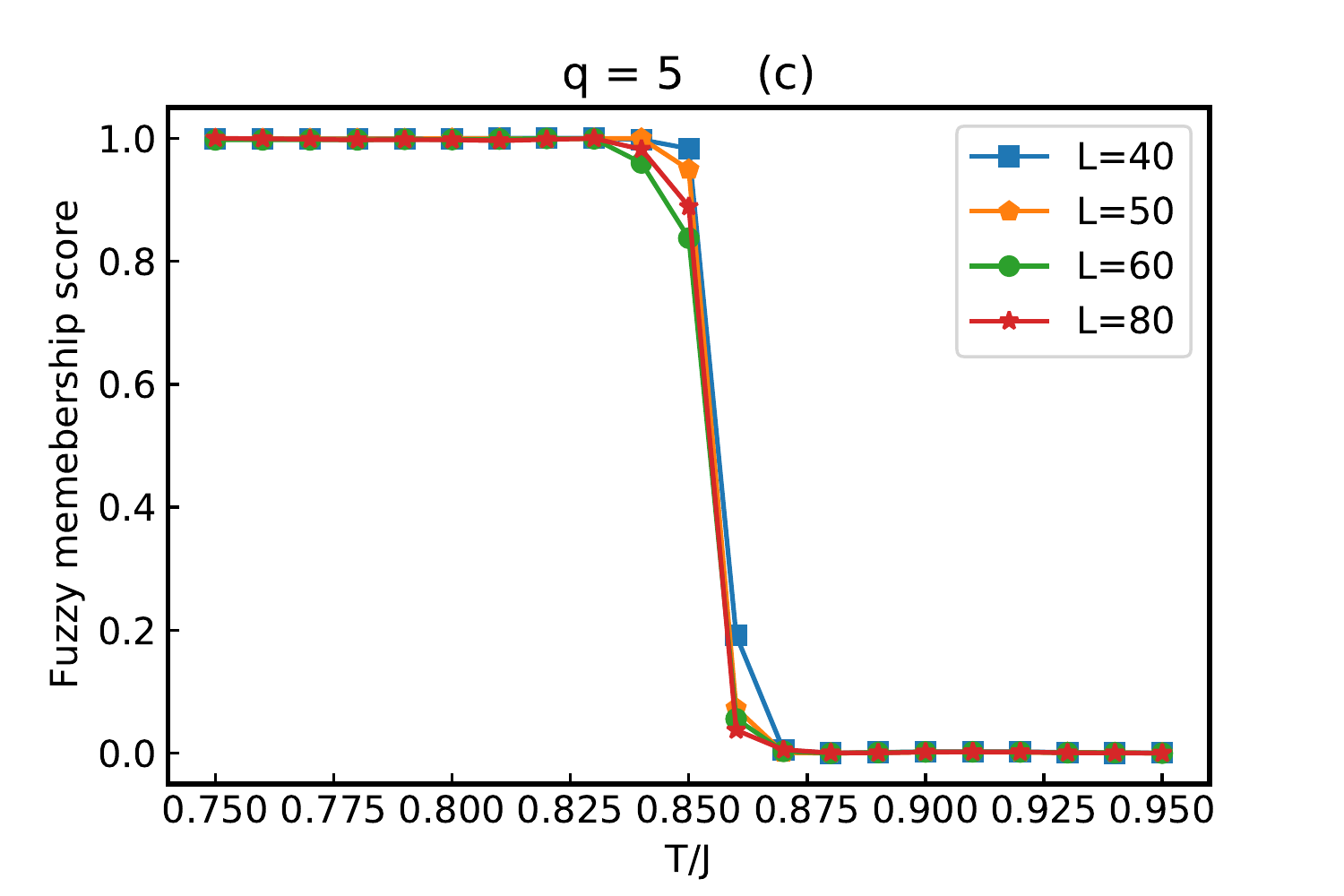}
	\caption{Fuzzy membership degree vector score as a function of temperature, for (a) $q=3$, (b) $q=4$, and (c) $q=5$; different data symbols refer to different linear sizes, $L$, as shown. 
	}
	\label{fig:mem_deg_tda}
\end{figure}

\begin{figure}[t]
	\centering
	\includegraphics[scale=0.44]{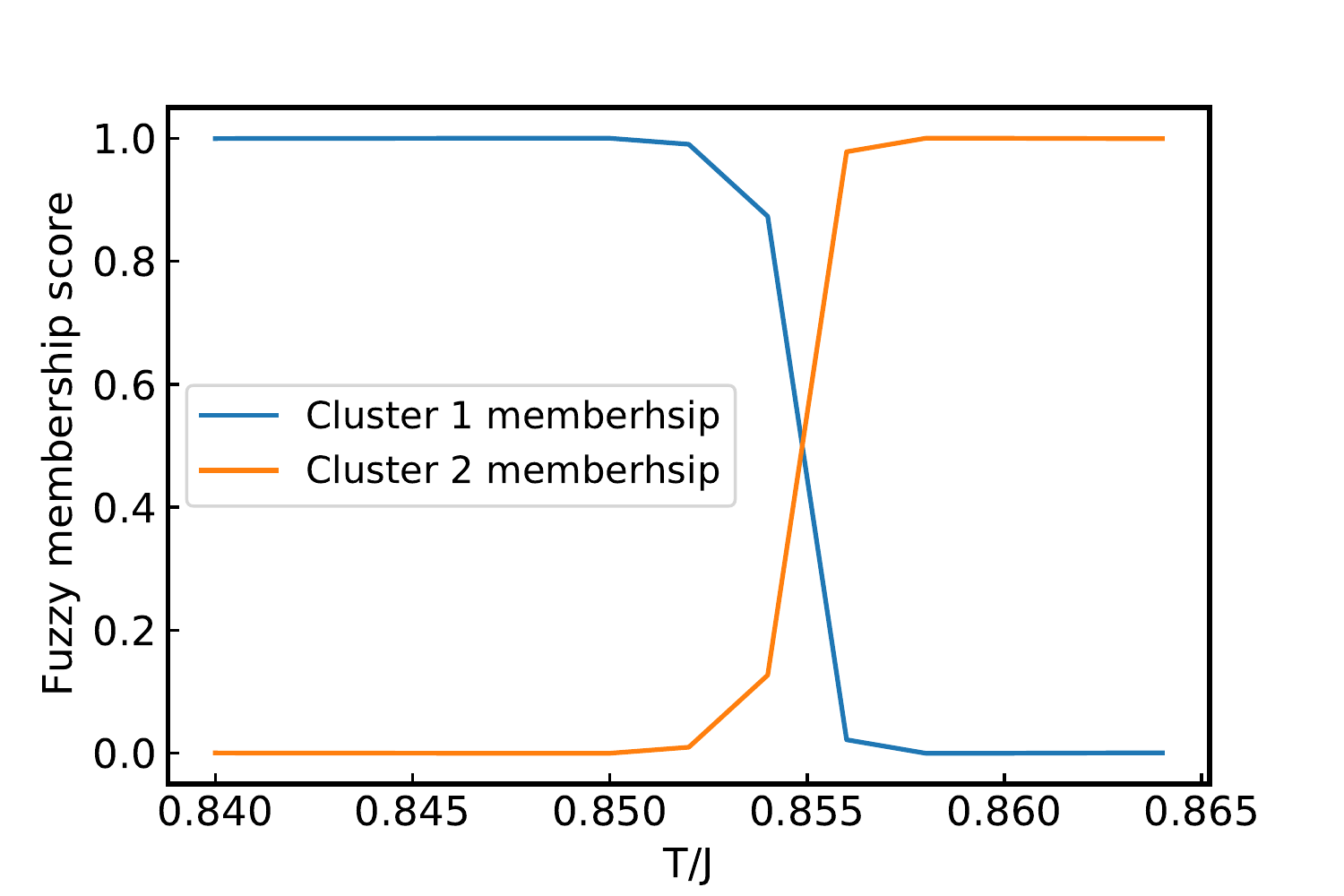}
	\caption{Fuzzy membership degree vector scores as a function of temperature, for the two clusters, for $q=5$ and $L=40$. Our estimate of the critical temperature is given by the intersection point of the two curves.
	}
	\label{fig:comp_mem}
\end{figure}
Nonetheless, as an additional test, we investigate the fuzzy membership degree curves for the dataset, Eq.\,\eqref{eq:member_list}, from which we expect that data within the same phase should have similar scores.
Figure \ref{fig:mem_deg_tda} displays the membership scores as functions of temperature, for $q=3$, 4, and 5.
In each case, the membership scores are 1 below the expected critical temperature, and drop sharply to zero above at higher temperatures. From the membership score functions we extract our estimate of $T_c$ as follows: within fuzzy clustering, the total membership of a point (i.e. a temperature in our case) amounts to 1 and has to be {\it partitioned} into partial membership values to each cluster. Therefore, when the number of clusters is two, the fuzzy membership functions will have the complementary trend shown in Fig. \ref{fig:comp_mem}. The intersection point of the two curves is our estimate of $T_c$. Moreover, the error bar associated to such an estimate is the width of the grid on which we sample $T/J$, (which in our case is $0.01$). Clearly, the error is systematically improvable by using a finer temperature grid.



\begin{figure}[t]
	\centering
	\includegraphics[scale=0.62]{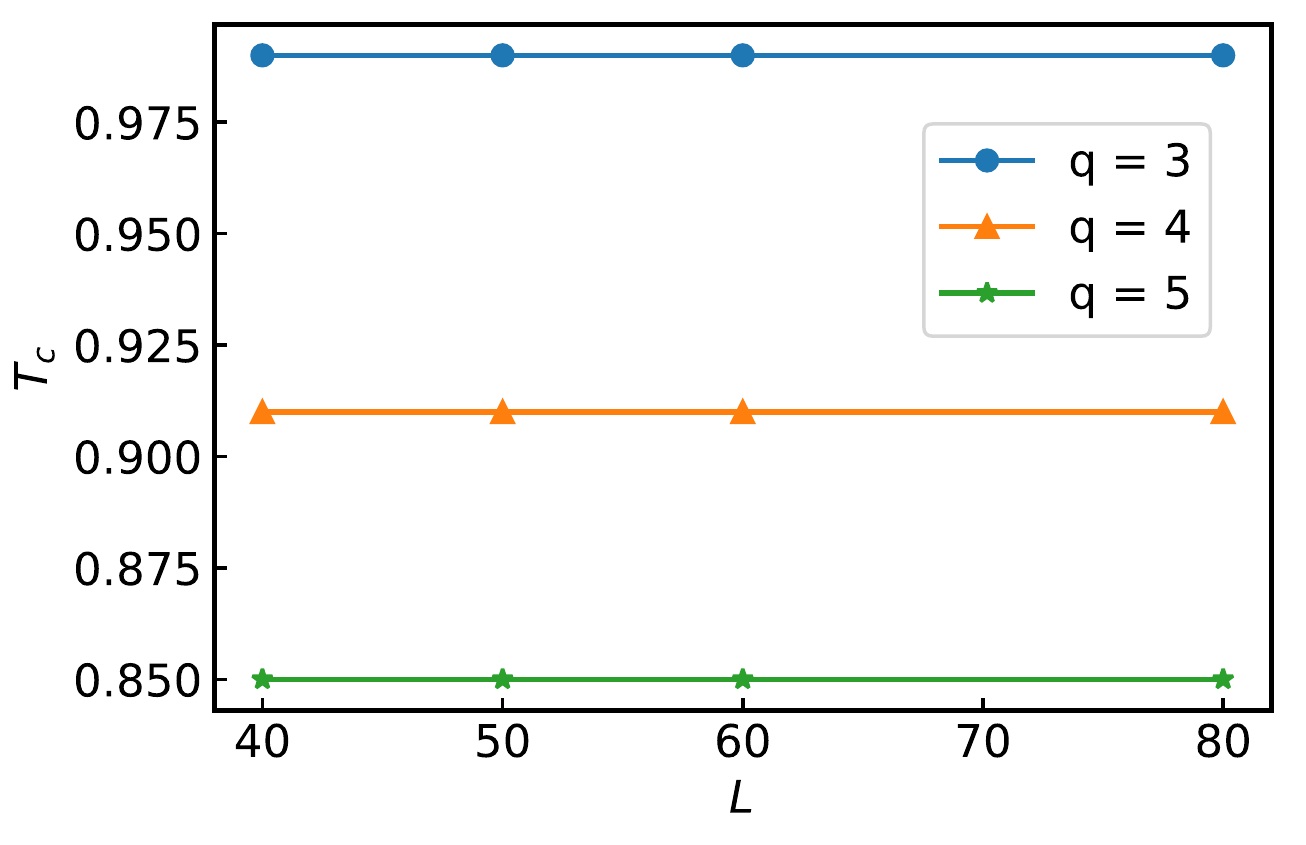} 
	\caption{Critical temperature of the $q$-state Potts model as a function of the lattice size for $q=3$, $q=4$ and $q=5$. As it can be noticed, the TDA technique does not require finite size scaling, as the transition temperature are fully independent from the lattice size.}
	\label{fig:crit_tda}
\end{figure}

For each $q$, we extract the critical temperature for each lattice size, and plot the resulting $T_c(L)$ in Fig.\,\ref{fig:crit_tda}.
We see that there is hardly any size-dependence of the estimated $T_c$'s, in marked contrast  with Fig.\,\ref{fig:scaling}.
The inescapable conclusion is that TDA drastically reduces finite-size effects in the determination of the critical temperature. 
On the other hand, in view of the almost step-like character of the data in Fig.\,\ref{fig:mem_deg_tda}, it does not track the order parameter; hence, one cannot resort to Eq.\,\eqref{eq:exponents} to obtain reliable critical exponents through this method.

\section{Conclusion} \label{sec:concl}

In this work we have tested the efficiency of a series of unsupervised Machine Learning (UML) algorithms in the study of classical phase transitions. 
To this end we have focused on the square lattice $q$-state Potts model since it provides quite stringent tests. 
Indeed, by varying $q$ one spans not only different universality classes, but also the nature (i.e.\ continuous or discontinuous) of the transitions.
In addition, probing how finite-size effects  manifest themselves in the different UML algorithms is a crucial methodological attribute of efficiency. 

More specifically, we have used Monte Carlo data to feed not just classical ML techniques, such as PCA and $k$-means, but additionally some new methods originating from Algebraic Topology, such as UMAP and TDA\,\cite{tirelli2021}.
We have examined the Potts model with $q=3$, 4 and 5, and all techniques yielded very accurate critical temperatures, $T_c(q)$, for the phase transition of the system.
In the case of PCA, we have used the finite-size effects (present in most techniques here) to our advantage by calculating critical exponents.  
Though not as accurate as the critical temperatures, our calculated exponents are in good agreement with the know exact values; 
most importantly, through them we were able to 
indicate differences 
between continuous and discontinuous transitions. 

Also, the dimensionality reduction techniques used, namely the PCA and UMAP, provide clues about the symmetry of the Hamiltonian, with the number of clusters at low-temperature being equal to the degenerate states of the model, i.e. those which break the Hamiltonian symmetry.

An important outcome of our comparative analysis is that linear techniques are more susceptible to finite size effects than non-linear ones, while the topologically inspired techniques are less sensitive to the system size.
In other words, despite demanding a more complex implementation the TDA approach leads to more accurate and reliable results, even for reasonably small system sizes.
Therefore, through further testing of these methodologies we expect this class of unsupervised machine learning approaches will become useful tools in the study of more challenging problems, for which standard techniques are unable to provide conclusive responses, both for classical and quantum systems.

\section*{ACKNOWLEDGMENTS}
A.T.~acknowledges financial support from the MIUR Progetti di Ricerca di Rilevante Interesse Nazionale (PRIN) Bando 2017 - grant 2017BZPKSZ.
A.T.~and N.C.C.~acknowledge CINECA for awarding them access to the Marconi100 supercomputer, through the ISCRA framework, within the projects AI-H-QMC - HP10BGJH1X, and IsB23 (ISCRA-HP10BF65I0).
N.C.C.~acknowledges financial support from the Brazilian Agency National Council for Scientific and Technological Development (CNPq), grant number 313065/2021-7.
D.O.C., L.A.O., J.P.L., N.C.C., and R.R.d.S.~are grateful to the Brazilian Agencies CNPq, National Council for the Improvement of Higher Education (CAPES), FAPERJ, and FAPEPI for partially funding this project.

\bibliography{references}
 
\end{document}